\documentclass[lettersize,journal]{IEEEtran}
\usepackage[fleqn]{amsmath}
\usepackage{amsfonts,amssymb}
\allowdisplaybreaks[4]
\usepackage{array}
\usepackage[caption=false,font=footnotesize,labelfont=sf,textfont=sf]{subfig}
\usepackage{textcomp}
\usepackage{stfloats}
\usepackage{url}
\usepackage{verbatim}
\usepackage{graphicx}
\usepackage{cite}
\usepackage[T1]{fontenc}
\usepackage[utf8]{inputenc}   

\usepackage{multirow}
\usepackage{float}
\AtBeginDocument{%
  \setlength{\abovedisplayskip}{4.5pt plus 2pt minus 1pt}%
  \setlength{\belowdisplayskip}{\abovedisplayskip}%
  \setlength{\belowdisplayshortskip}{\abovedisplayskip}%
}
\usepackage{eqparbox}
\usepackage{mdwmath}
\usepackage{mdwtab}

\usepackage{algorithm}
\usepackage{algpseudocode}  
\usepackage{csquotes}
\usepackage{enumitem}
\usepackage{ifthen}
\usepackage{nomencl}
\usepackage{xfrac}            
\usepackage{nicefrac}      

\usepackage{xcolor}
\usepackage{upgreek}

\newcommand{\beq}{\begin{equation}}
\newcommand{\eeq}{\end{equation}}
\newcommand{\beqn}{\begin{eqnarray}}
\newcommand{\eeqn}{\end{eqnarray}}
\newcommand{\beqno}{\begin{eqnarray*}}
\newcommand{\eeqno}{\end{eqnarray*}}
\newcommand{\bma}{\begin{displaymath}}
\newcommand{\ema}{\end{displaymath}}
\newcommand{\bnu}{\begin{enumerate}}
\newcommand{\enu}{\end{enumerate}}
\newcommand{\bce}{\begin{center}}
\newcommand{\ece}{\end{center}}
\newcommand{\btb}{\begin{tabular}}
\newcommand{\etb}{\end{tabular}}

\usepackage{pifont}

\usepackage{amsthm}
\theoremstyle{plain}

\theoremstyle{definition}

\usepackage[hidelinks]{hyperref}

\makeatletter
\let\TSGoriginalmakecaption\@makecaption
\def\TSGfiguretype{figure}
\long\def\@makecaption#1#2{%
  \ifx\@captype\TSGfiguretype
    \@IEEEfigurecaptionsepspace
    \parbox[t]{\hsize}{\normalfont\footnotesize\centering
      #1.\nobreakspace\nobreakspace #2\par}%
  \else
    \TSGoriginalmakecaption{#1}{#2}%
  \fi}
\makeatother

\begin{document}

\title{Checkpoint-Aware Coordination of LLM Training Loads for Power Grid Flexibility}

\author{Jiaming Cheng,~\IEEEmembership{Student Member,~IEEE,}
        and Duong Tung Nguyen,~\IEEEmembership{Member,~IEEE}
\thanks{The authors are with the Ira A. Fulton Schools of Engineering, Arizona State University, Tempe, AZ, USA. Email: \{jiaming,duongnt\}@asu.edu. }}

\maketitle


\begin{abstract}
Pausing large language model training can reduce demand during a
constrained period. The grid must also serve the load when jobs resume.
We distinguish executed from saved work to schedule training,
checkpoint writes, resource release, and recovery jointly with grid
dispatch and payments to data centers. Column generation uses
shortest-path pricing to bound the optimal cost. On a 73-bus network,
a plan chosen without line limits fails peak-load redispatch.
Joint coordination serves the same jobs within the original ratings
and lowers grid cost plus payments relative to the agreed baseline in
all 40 realized-input cases. Across these cases, choosing the first
write and using a fixed two-hour interval of training work thereafter
retains at least 90.37\% of the optimal net savings available with
per-write timing.
In a three-bus case with equal write counts, choosing every write reduces
local generation under tight deadlines and uneven import windows.
Forecast-based planning raises realized cost on the
tested peak day even with fast-unit recommitment and startup-inclusive
accounting.
\end{abstract}


\section{Introduction}
\label{sec:introduction}

U.S. data-center electricity demand is projected to grow through
2030~\cite{smith2026datacenter}. A field trial reduced cluster power for
several hours by adjusting workloads~\cite{colangelo2025phoenix}.
Large language model (LLM) training can shift demand by pausing jobs
and finishing their work later.

A grid coordinator must decide how much training to defer and when
jobs can resume within grid limits. Holding keeps graphics processing
units (GPUs) reserved. Releasing them frees capacity, and the job must
reload a checkpoint or repeat work to resume. Writes and restores use
time, power, and shared storage input/output
(I/O)~\cite{mohan2021checkfreq,lian2025universal}. Concurrent restarts compete for storage and can overload a line.
The coordinator must serve recovery while each campus meets deadlines
and accounts for writing, waiting, and repeated work.

Prior work coordinates workload shifts through dispatch or
prices~\cite{tsiligkaridis2025distributed,zhang2024aggregator}.
Checkpoint-aware dispatch takes execution into account by scheduling
waits and GPU modes at preset progress boundaries~\cite{tong2026execution}
or by using checkpoint alerts to manage uncertain write
times~\cite{liu2026privacycheckpoint}.
Both checkpoint-aware approaches take save points or candidate write
times as inputs. This leaves the choice of where in training to save
before GPU release outside the dispatch problem. Moving a write can
change the work a job must repeat and when the grid serves recovery.

We choose complete job trajectories jointly with grid dispatch and
campus payments. Paths record executed and saved work, power, GPU use,
and I/O. Shared campus resources and direct-current optimal power flow
(DCOPF) constraints couple the paths. Payments cover cost increases over
the agreed schedule.
\begin{itemize}[nosep]
\item \textbf{Modeling.} We represent holding, GPU release, restoration,
and recomputation under common deadlines and resource limits. The model
can choose each write or optimize the first write and fix the interval
in training work for later writes.
\item \textbf{Solution approach.} Grid-aware checkpoint trajectory
generation (GACTG) prices job trajectories by shortest paths using dual
prices for power, shared resources, and compensation. It returns a
feasible schedule and a lower bound on optimal cost.
\item \textbf{Numerical results.} Fixed-load redispatch shows how omitting
line limits can yield an infeasible plan. Matched comparisons distinguish
savings from rescheduling training, releasing GPUs, and choosing write
times. The first-write mode retains most savings on the reference days.
With equal write counts, choosing each write reduces local generation
when uneven import windows leave little time to wait.
\end{itemize}

\section{Related Work}
\label{sec:related_work}

\noindent \textbf{\textit{Grid coordination and flexibility procurement.}}
Workload shifts can reduce electricity costs or
emissions~\cite{liu2011greening,hanafy2023carbonscaler,radovanovic2023carbon}.
Trace, field, and interconnection studies quantify flexibility and
capacity impacts~\cite{colangelo2025phoenix,caprara2026flexibility,acun2026flexdc,norris2025rethinking}.
Dispatch and expansion models represent flexible
demand~\cite{tsiligkaridis2025distributed,tsg2021aggregatedIDC}, and
procurement studies design incentives~\cite{zhang2024aggregator,tsg2021incentiveIDC}.
These models leave checkpoint and recovery loads implicit.

\noindent \textbf{\textit{Training execution and checkpoint recovery.}}
Parallel training systems configure GPU allocations~\cite{zheng2022alpa}.
Checkpointing balances write overhead and lost work~\cite{mohan2021checkfreq}
and supports repartitioning~\cite{lian2025universal} and preemption~\cite{thorpe2023bamboo}.
Measurements characterize training power~\cite{vercellino2026genai,choukse2025power},
and host-memory staging overlaps training with durable writes~\cite{hu2024acmetrace,wan2025bytecheckpoint}.
Timing writes and concurrent restores within transmission limits lies
outside these studies.

\noindent \textbf{\textit{Checkpoint-aware power-system coordination.}}
Checkpoint-aware dispatch chooses waits and GPU modes at fixed
progress boundaries~\cite{tong2026execution} or uses candidate write windows
from the campus~\cite{liu2026privacycheckpoint}. Batteries buffer
checkpoint power changes~\cite{lu2026battery}.
Other models represent interruption as virtual
storage~\cite{liu2027interruptible} or choose a site-level repeat interval
using average power and expected failure loss~\cite{zhang2026carbon}.
These formulations leave individual save-point choices outside
network dispatch.

We coordinate each job's save points, release, and recovery with network
dispatch and campus compensation. Job trajectories account for write
and recovery loads, shared resources, and recomputation.

\section{System Model}
\label{sec:system}

\begin{figure}[!t]
\centering
\includegraphics[width=\columnwidth]{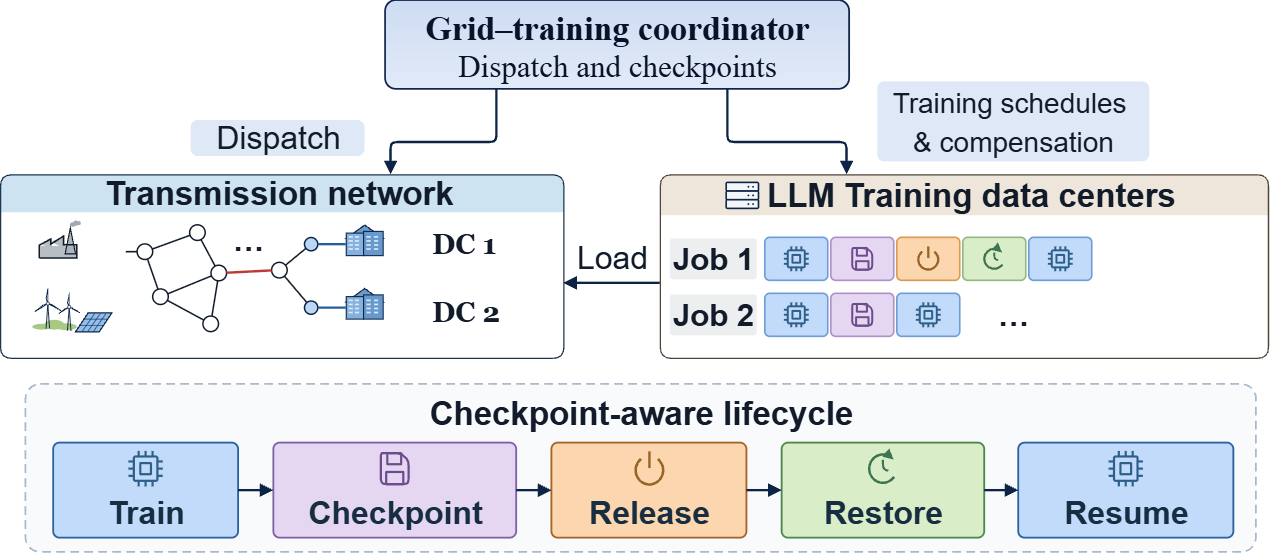}
\caption{Joint dispatch, training paths, and campus payments.
The selected paths determine the loads at the campus connection buses.}
\label{fig:system_model}
\end{figure}

The coordinator plans dispatch and training in slots
$\mathcal{T}=\{0,\ldots,T-1\}$. Each slot lasts $\Delta$ hours.
Slot boundaries are $\mathcal{T}^{+}=\{0,\ldots,T\}$.
Power is in MW, energy in MWh, and storage bandwidth in GB/s.
Fig.~\ref{fig:system_model} illustrates the coordination between the power
grid and participating artificial intelligence (AI) data centers, including
the checkpoint-aware training lifecycle.

\subsection{Power Network and Resources}
\label{subsec:power_network}
Let $\mathcal{N},\mathcal{E},\mathcal{G},\mathcal{R},\mathcal{I}$ denote buses,
lines, conventional generators, renewables, and data centers.
Resources $g$, $r$, and $i$ connect at $n(g)$, $n(r)$, and $n(i)$.
Bus $n_0$ is the angle reference. Generator output $p_{g,t}$ has limits
$\underline P_g,\overline P_g$ and fixed binary commitment $u_{g,t}$.
Its convex piecewise-linear cost $C_{g,t}(p_{g,t})$ in \$/h includes
no-load cost online and is zero offline. Ramp limit $R_g$ is in MW/slot.
Startup and shutdown limit output to $S_g=\max\{\underline P_g,R_g\}$.
Renewable availability $\overline P_{r,t}^{\mathrm{ren}}$ is dispatched as
$p_{r,t}^{\mathrm{ren}}$, with curtailment
\begin{equation}
p_{r,t}^{\mathrm{curt}}
=
\overline P_{r,t}^{\mathrm{ren}}
-p_{r,t}^{\mathrm{ren}}.
\label{eq:renewable_curtailment}
\end{equation}
Curtailment incurs a penalty $c_r^{\mathrm{curt}}\ge0$ in \$/MWh.
The lossless direct-current (DC) approximation represents active-power
dispatch, ramping, and transmission limits~\cite{tsiligkaridis2025distributed}.
It excludes reactive power, voltage limits, and frequency dynamics.
Line $e=(n',n)$ has susceptance $b_e$ (including any transformer tap)
and rating $\overline F_e$. Its signed flow is $f_{e,t}$.
Bus angles are $\theta_{n,t}$ and fixed demand is $P_{n,t}^{\mathrm{fix}}$.

\subsection{Training Lifecycle}
\label{subsec:training_lifecycle}

\noindent \textbf{\textit{Requirements and modes.}}
Each job belongs to one campus job set $\mathcal{J}_i$, with
$\mathcal{J}=\bigcup_i\mathcal{J}_i$. Job $j$ has arrival boundary
$A_j\in\mathbb Z$, work $W_j$, and deadline $D_j\in\mathcal{T}^{+}$.
Work counts optimizer steps or a fixed normalization. Model, data,
batch size, sequence length, and optimizer are fixed.
Each feasible mode $m\in\mathcal M_j$ (a finite set) specifies
gang size $G_{jm}$, rate $\varrho_{jm}$ in work units/hour, and
incremental information technology (IT) power $P_{jm}^{\mathrm{tr}}$,
subject to memory and parallelism requirements~\cite{zheng2022alpa}.
GPUs are interchangeable within a pool. Mode changes release the
gang before compatible restoration or initialization.

The nominal work increment is $\Delta x_{jm}=\varrho_{jm}\Delta>0$.
The finite grid $\mathcal{X}_j$ contains $0$, $W_j$, initial progress,
and all progress reachable by clipped increments in the horizon.
For every mode and policy, a training slot credits at most the remaining
work and reserves a full slot of GPUs, I/O, and power.

\noindent \textbf{\textit{Executed and saved progress.}}
Let $\mathcal{K}_j$ list supported checkpoint formats and
$\overline U_j$ be the maximum unsaved work in the same units as $W_j$.
At an operation boundary, a job's state and progress satisfy
\begingroup
\allowdisplaybreaks[0]
\begin{subequations}\label{eq:lifecycle_state_group}
\begin{align}
s=(t,x,q,v,m,k),~ t\in\mathcal{T}^{+}, \label{eq:lifecycle_state}\\
v\in\{\mathrm{new},\mathrm{active},\mathrm{off},\mathrm{done}\},
  \label{eq:lifecycle_status}\\
m\in\mathcal{M}_j\cup\{\varnothing\},~
k\in\mathcal{K}_j\cup\{\varnothing\}, \label{eq:lifecycle_domains}\\
x,q\in\mathcal{X}_j, \label{eq:lifecycle_progress_grid}\\
0\le q\le x\le W_j, \label{eq:training_progress}\\
U_j(s):=x-q\le\overline U_j. \label{eq:unprotected_progress}
\end{align}
\end{subequations}
\endgroup
Here $x$ is executed progress and $q$ is durably saved progress.
Active jobs have $m\in\mathcal M_j$, with $m=\varnothing$ otherwise.
Checkpoint format $k=\varnothing$ implies $q=0$, which can also be saved
in a valid checkpoint. Recovery uses the latest completed
checkpoint~\cite{mohan2021checkfreq,lian2025universal}.

All policies start job $j$ at the same admissible state
$s_j^0=(0,x_j^0,q_j^0,v_j^0,m_j^0,k_j^0)$.
New jobs have $x=q=0$ and no checkpoint.
Initially active jobs have $A_j\le0$. Off jobs have $x=q$.

\noindent \textbf{\textit{Training, holding, and writing.}}
Training requires an active job. Its clipped work
increment and state update are
\begingroup
\setlength{\abovedisplayskip}{2pt}
\begin{subequations}\label{eq:training_transition_group}
\begin{align}
\gamma_{jm}(x)&=\min\{\Delta x_{jm},W_j-x\},~ 0\le x<W_j,
  \label{eq:productive_gain}\\
(t,x,q)&\longmapsto(t+1,x+\gamma_{jm}(x),q),~ t+1\le D_j.
  \label{eq:productive_transition}
\end{align}
\end{subequations}
\endgroup
An active job can hold for one slot at fixed progress and mode.
It retains $G_{jm}$ GPUs, draws $P_{jm}^{\mathrm{hold}}\ge0$, and can
resume training without reloading state.

An active job writes synchronously for $\tau_{jm}^{\mathrm{ck}}\ge1$ slots.
In the write update, primes denote end-state values.
\begin{align}\label{eq:checkpoint_transition}
t'&=t+\tau_{jm}^{\mathrm{ck}},~ x'=x, ~ q'=x,~ k'=k_{jm}^{\mathrm{ck}}.
\end{align}
Here $k_{jm}^{\mathrm{ck}}$ is the format saved in mode $m$.
The job stays active and uses its write profile, including when $x=q$.

\noindent \textbf{\textit{Release, recovery, and completion.}}
A graceful stop requires an active job with a valid checkpoint and
$x=q$. An abrupt stop of an active job discards $x-q$ by setting $x'=q$.
Both enter the off state and release the gang after
$\tau_{jm}^{\mathrm{sd}}\ge0$ shutdown slots. A zero-slot shutdown
can connect distinct states at the same time boundary.

New and off jobs can wait. With $k\ne\varnothing$ and binary
compatibility $\chi_j(k,m')=1$, an off job restores to active mode $m'$
with $x'=q$. Training then repeats any work discarded at release.
Reading, repartitioning, and recovery use
$\tau_{jkm'}^{\mathrm{rs}}\ge1$ slots~\cite{lian2025universal,thorpe2023bamboo}.
New and off jobs can also initialize in mode $m'$ after
$\tau_{jm'}^{\mathrm{init}}\ge1$ slots, setting $x'=q'=0$ and
$k'=\varnothing$.

Waiting uses no job resources for one slot. Other operations start at
or after $A_j$. Each runs to completion by $D_j$ and reserves its
profiles in slots $t,\ldots,t'-1$. Endpoints obey
\eqref{eq:lifecycle_state_group} and preserve components unless specified.
Finalization starts active at $x=W_j$, takes $\tau_{jm}^{\mathrm{fin}}\ge1$
slots, saves format $k_{jm}^{\mathrm{ck}}$, and releases the gang. Then
\begin{align}\label{eq:lifecycle_terminal}
v =\mathrm{done},~ x=q=W_j, m =\varnothing,~ k\ne\varnothing,~ t\le D_j.
\end{align}
Each training arc $l$ adds $\gamma_{j,m(l)}(x_l)$, where $x_l$ and $m(l)$
are its source progress and mode. Abrupt stops lose $x-q$, and resets
from off lose the retained $q$. Other arcs leave $x$ unchanged.
Summing these changes from $x_j^0$ to terminal progress $W_j$ gives
\begin{equation}
x_j^0+\sum_{l\in\mathcal{L}_{jh}^{\mathrm{tr}}}
\gamma_{j,m(l)}(x_l)-\sum_{l\in h}\omega_{jl}=W_j,
\label{eq:lifecycle_work_accounting}
\end{equation}
where $\mathcal{L}_{jh}^{\mathrm{tr}}$ collects training arcs in path $h$
and $\omega_{jl}$ is discarded progress on abrupt-stop and reset arcs
and zero on all other arcs.

\subsection{Trajectories and Shared Resources}
\label{subsec:trajectory_interface}
For job $j$, graph $\mathcal{Q}_j=(\mathcal{S}_j,\mathcal{L}_j)$ has
admissible states as nodes and complete legal operations as arcs.
The rank $(t,\mathbf 1_{\{v=\mathrm{off}\}})$ increases lexicographically.
Positive durations advance time and zero-duration shutdown goes from
active to off. Thus the graph is acyclic. The paths $\mathcal H_j$ from
$s_j^0$ to \eqref{eq:lifecycle_terminal} correspond to legal sequences
because each arc joins admissible states and passes its endpoint to
the next arc. The selected path is $h_j$.

For arc $l$ of job $j$, $G_{jl,t}$, $B_{jl,t}$, and
$p_{jl,t}^{\mathrm{IT}}$ give GPU occupancy, storage bandwidth, and
incremental IT power in slot $t$. These nonnegative profiles are zero
outside the operation's slots. We sum them along each path.
\begin{align}\label{eq:trajectory_profiles}
G_{jh,t}=\sum_{l\in h}G_{jl,t},~ B_{jh,t}=\sum_{l\in h}B_{jl,t}, ~ p_{jh,t}^{\mathrm{IT}}=\sum_{l\in h}p_{jl,t}^{\mathrm{IT}}.
\end{align}
Campus limits $G_i^{\max},B_i^{\max},P_i^{\max}$ apply to GPU use,
storage bandwidth, and facility power. Writes, restores, and finalization
share I/O capacity. We assume enough persistent storage for all
retained checkpoints and in-progress replacements.
The selected IT profile and facility demand are
\begin{subequations}\label{eq:facility_power_group}
\begin{align}
p_{j,t}^{\mathrm{IT}}&=p_{jh_j,t}^{\mathrm{IT}},\\
P_{i,t}^{\mathrm{DC}}
&=P_{i,t}^{\mathrm{idle}}+\eta_i\sum_{j\in\mathcal{J}_i}p_{j,t}^{\mathrm{IT}}.
\label{eq:dc_aggregate_power}
\end{align}
\end{subequations}
Idle power $P_{i,t}^{\mathrm{idle}}\ge0$ covers hardware and auxiliaries.
Operations add power above it, zero for off-GPU waits. Each additional
IT MW adds $\eta_i\ge1$ facility MW. Power usage effectiveness measures
total facility energy divided by IT energy~\cite{doe2024datacenter}.
Facility energy per slot is $P_{i,t}^{\mathrm{DC}}\Delta$.

\subsection{Baselines and Private Costs}
\label{subsec:contractual_baselines}
The agreed paths $h_j^{\mathrm{base}}\in\mathcal H_j$ jointly meet
resource and network limits~\cite{zhang2024aggregator,tsg2021incentiveIDC}.
Tariff $\rho_{i,t}$ is in \$/MWh. Additive fees $c_{jh}^{\mathrm{op}}$
cover operations, nontraining GPU reservations, and lost work.
Arc endpoints determine discarded progress. The single-mode tests
charge each lost unit for its known GPU hours. Campus payments use
the total private-cost change.
\begin{subequations}\label{eq:private_cost_group}
\begin{align}
c_{jh}&=\sum_{t\in\mathcal{T}}\rho_{i,t}\eta_i
 p_{jh,t}^{\mathrm{IT}}\Delta+c_{jh}^{\mathrm{op}}, \label{eq:job_private_cost}\\
\Delta c_{jh}&=c_{jh}-c_{j,h_j^{\mathrm{base}}}, \label{eq:job_incremental_cost}\\
\Delta C_i&=\sum_{j\in\mathcal{J}_i}\Delta c_{j,h_j}, \label{eq:campus_incremental_cost}\\
M_i&=\max\{0,\Delta C_i\}. \label{eq:minimum_reimbursement}
\end{align}
\end{subequations}
Fixed costs cancel; demand charges and other nonadditive costs are excluded.
For baseline private cost $C_i^{\mathrm{base}}$, the reimbursed campus cost obeys
\begin{equation}
C_i^{\mathrm{base}}+\Delta C_i-M_i
=C_i^{\mathrm{base}}+\min\{0,\Delta C_i\}
\le C_i^{\mathrm{base}}.
\label{eq:contract_participation}
\end{equation}
Participation does not increase modeled private cost under accurate inputs.
This condition does not establish truthful reporting or incentive compatibility.

\noindent \textbf{\textit{Operational implementation.}}
Under the assumed voluntary agreements, each data center
provides supported training modes, checkpoint and recovery
profiles, resource limits, and an agreed baseline schedule.
The grid-side coordinator jointly selects training trajectories
and network-constrained generation dispatch. The data-center
scheduler executes the selected checkpoint, GPU release,
and restoration operations through its training runtime.
Updated job states and resource availability could support
subsequent coordination rounds.

\section{Problem Formulation}
\label{sec:formulation}
\label{subsec:coordination_scope}
Binary $z_{jh}$ selects trajectory $h$ for job $j$. The coordinator
chooses trajectories, dispatch, angles, line flows, and payments to
minimize grid operating cost plus compensation. The trajectories set
campus demand in \eqref{eq:selected_dc_power}, which enters the multi-period
DCOPF in \eqref{eq:generator_limits}--\eqref{eq:nodal_balance}~\cite{stott2009dc}.
Load, renewable supply, and commitment $u_{g,t}$ are given. The reference
study uses baseline-demand unit commitment (UC) and realized profiles.
\S\ref{subsec:forecast} evaluates day-ahead plans.

At bus $n$, $\mathcal{G}_n,\mathcal{R}_n,\mathcal{I}_n$ collect local
resources and $\mathcal{E}_n^{\mathrm{in}},\mathcal{E}_n^{\mathrm{out}}$
collect incoming and outgoing lines. The model is
\begingroup
\setlength{\jot}{1pt}
\begin{subequations}\label{eq:joint_coordination}
\begin{align}
&\mathcal P:~\min~
 \Delta\!\Big[\sum_{g,t} C_{g,t}(p_{g,t})
 +\sum_{r,t} c_r^{\mathrm{curt}}p_{r,t}^{\mathrm{curt}}\Big]+\sum_i M_i,
 \label{eq:grid_operating_cost}\\
&\mathrm{s.t.}~\sum_{h\in\mathcal H_j}z_{jh}=1,~ \forall j,
 \label{eq:trajectory_selection}\\
&\sum_{\substack{j\in\mathcal J_i\\h\in\mathcal H_j}}G_{jh,t}z_{jh}
 \le G_i^{\max},~ \forall i,t,
 \label{eq:dc_gpu_capacity}\\
&\sum_{\substack{j\in\mathcal J_i\\h\in\mathcal H_j}}B_{jh,t}z_{jh}
 \le B_i^{\max},~ \forall i,t,
 \label{eq:dc_storage_capacity}\\
&P_{i,t}^{\mathrm{DC}}=P_{i,t}^{\mathrm{idle}}
 +\eta_i\!\sum_{\substack{j\in\mathcal J_i\\h\in\mathcal H_j}}
 p_{jh,t}^{\mathrm{IT}}z_{jh},~ \forall i,t,
 \label{eq:selected_dc_power}\\
&0\le P_{i,t}^{\mathrm{DC}}\le P_i^{\max},~ \forall i,t,
 \label{eq:dc_interconnection}\\
&\underline P_gu_{g,t}\le p_{g,t}\le\overline P_gu_{g,t},~ \forall g,t,
 \label{eq:generator_limits}\\
&0\le p_{r,t}^{\mathrm{ren}}\le\overline P_{r,t}^{\mathrm{ren}},~ \forall r,t,
 \label{eq:renewable_limits}\\
&p_{g,t}-p_{g,t-1}\le R_gu_{g,t-1}+S_g(1-u_{g,t-1}),\ \forall g,t\ge1,
 \label{eq:ramp_up_forward}\\
&p_{g,t-1}-p_{g,t}\le R_gu_{g,t}+S_g(1-u_{g,t}),\ \forall g,t\ge1,
 \label{eq:ramp_down}\\
&f_{e,t}=b_e(\theta_{n',t}-\theta_{n,t}),~ \forall e=(n',n),t,
 \label{eq:dc_line_flow}\\
&|f_{e,t}|\le\overline F_e,~ \forall e,t,
 \label{eq:line_capacity}\\
&\theta_{n_0,t}=0,~ \forall t,
 \label{eq:angle_reference}\\
&\sum_{g\in\mathcal G_n}p_{g,t}
 +\sum_{r\in\mathcal R_n}p_{r,t}^{\mathrm{ren}}
 +\sum_{e\in\mathcal E_n^{\mathrm{in}}}f_{e,t}\nonumber\\*
&~=P_{n,t}^{\mathrm{fix}}
 +\sum_{i\in\mathcal I_n}P_{i,t}^{\mathrm{DC}}
 +\sum_{e\in\mathcal E_n^{\mathrm{out}}}f_{e,t},~ \forall n,t,
 \label{eq:nodal_balance}\\
&M_i\ge0,~
 M_i\ge\!\sum_{\substack{j\in\mathcal J_i\\h\in\mathcal H_j}}\Delta c_{jh}z_{jh},
 ~ \forall i,
 \label{eq:compensation_epigraph}\\
&z_{jh}\in\{0,1\},~ \forall j,h.
 \label{eq:trajectory_binary}
\end{align}
\end{subequations}
\endgroup
\eqref{eq:trajectory_selection} gives each job one unit of weight over
its legal paths. \eqref{eq:dc_gpu_capacity} and
\eqref{eq:dc_storage_capacity} limit GPU occupancy and shared storage
bandwidth, respectively, at each campus and slot. \eqref{eq:selected_dc_power} adds
idle power and scales IT load for facility overhead.
\eqref{eq:dc_interconnection} bounds campus demand between zero and
the interconnection limit.
\eqref{eq:generator_limits} bounds online output and sets offline
output to zero. \eqref{eq:renewable_limits} bounds renewable dispatch
by zero and availability. \eqref{eq:ramp_up_forward} and
\eqref{eq:ramp_down} limit increases and decreases to $R_g$ between
online slots and cap startup and shutdown output at $S_g$. They
start at $t=1$, with no pre-horizon output constraint.
\eqref{eq:dc_line_flow} computes signed flows from angle differences.
\eqref{eq:line_capacity} enforces ratings in both directions, and
\eqref{eq:angle_reference} fixes the reference angle at zero.
\eqref{eq:nodal_balance} balances generation and imports against
fixed demand, campus load, and exports at each bus.
\eqref{eq:compensation_epigraph} requires nonnegative payments that
cover any campus private-cost increase. Binary choices
\eqref{eq:trajectory_binary} select exactly one path per job.
The objective \eqref{eq:grid_operating_cost} adds payments to dispatch
cost $\mathcal C_{\mathrm{grid}}$, yielding \eqref{eq:minimum_reimbursement}.
Dispatch cost includes generation and curtailment penalties. Startup cost is
fixed by commitment, and tariffs enter private costs. Piecewise-linear
costs make $\mathcal P$ a mixed-integer linear program (MILP).

\subsection{Value and Problem Structure}
\label{subsec:economic_interpretation}
Let $C_{\mathrm{grid}}^{\mathrm{base}}$ be the minimum grid operating cost
at the agreed baselines and $C_{\mathrm{grid}}^{\mathrm{coord}}$ the grid
cost under a coordinated plan. Define gross and net values by
\begin{subequations}\label{eq:coordination_value_group}
\begin{align}
C_{\mathrm{proc}}^{\mathrm{coord}}&=\sum_{i\in\mathcal{I}}M_i,\\
V^{\mathrm{gross}}&=C_{\mathrm{grid}}^{\mathrm{base}}
 -C_{\mathrm{grid}}^{\mathrm{coord}}, \label{eq:net_coordination_value}\\
V^{\mathrm{net}}&=V^{\mathrm{gross}}-C_{\mathrm{proc}}^{\mathrm{coord}}.
 \label{eq:net_value_definition}
\end{align}
\end{subequations}
Selecting the agreed paths with baseline dispatch and zero payments is
feasible, so the optimal objective $C^\star\le C_{\mathrm{grid}}^{\mathrm{base}}$.
Thus $V^{\mathrm{net},*}=C_{\mathrm{grid}}^{\mathrm{base}}-C^\star\ge0$.
For any lower bound $LB$ and feasible cost $UB$,
subtracting $LB\le C^\star\le UB$ from the baseline cost gives
\begin{equation}
C_{\mathrm{grid}}^{\mathrm{base}}-UB
\le V^{\mathrm{net},*}\le C_{\mathrm{grid}}^{\mathrm{base}}-LB.
\label{eq:coordination_value_bounds}
\end{equation}

\section{Solution Approach}
\label{sec:algorithm}

A direct solve of \eqref{eq:joint_coordination} has one binary variable
per path and can require exponentially many variables. The arc-flow
MILP includes every operation arc and job flow balance, with the same
integer and relaxed optimal values (Appendix~\ref{app:repro}).
Dantzig--Wolfe decomposition exploits local job feasibility by generating
paths per job and keeping campus capacity, network dispatch, and payments
in the master~\cite{desrosiers2005primer,barnhart1998branch}.

GACTG records executed and saved progress, GPU allocation, and checkpoint
format in each acyclic job graph. The state determines what work
survives release and which restores are compatible. Each arc carries
slot-level resource use and additive private cost. The campus payment
constraint gives one dual weight to the local job-cost changes. Nodal
and resource duals complete the arc weights. Pricing then finds an exact
shortest path over all legal trajectories for each job. The restricted
master objective and minimum reduced costs lower-bound the full linear
programming (LP) objective. Replay and redispatch verify the recovered
integer schedule.

\subsection{Master Problem and Prices}
\label{subsec:gactg_master}
\label{subsec:gactg_pricing}

The restricted master problem (RMP) uses $z_{jh}\ge0$ and a path pool
$\widehat{\mathcal H}_j$, initialized with the agreed paths.
These paths ensure initial feasibility. Each solve updates path
weights, dispatch, and duals.
Substitute \eqref{eq:selected_dc_power} into the capacity and balance
constraints. The lower bound $P_{i,t}^{\mathrm{DC}}\ge0$ is redundant
by nonnegative profiles and idle power.
In the Lagrangian, $\alpha_{i,t},\beta_{i,t},\psi_{i,t},\kappa_i\ge0$
multiply GPU use, I/O, and power minus their limits, and the private
cost increase minus payment, respectively. Use $-\sigma_j$ on
$\sum_h z_{jh}-1$ and $\Delta\lambda_{n,t}$ on the right-hand side
minus the left-hand side of \eqref{eq:nodal_balance}.
Collecting the coefficient of $z_{jh}$ gives
\begin{equation}
\begin{aligned}
\bar c_{jh}={}&-\sigma_j+\kappa_i\Delta c_{jh}
+\sum_{t\in\mathcal{T}}\big[\alpha_{i,t}G_{jh,t}+\beta_{i,t}B_{jh,t}\\
&+(\psi_{i,t}+\Delta\lambda_{n(i),t})\eta_ip_{jh,t}^{\mathrm{IT}}\big].
\end{aligned}
\label{eq:gactg_reduced_cost}
\end{equation}
Here $\lambda_{n,t}$ is the locational marginal price (LMP) in \$/MWh.
The coefficient of $M_i$ is $1-\kappa_i$. Since $M_i\ge0$,
dual feasibility requires $\kappa_i\le1$. Complementary slackness gives $\kappa_i=1$ for $M_i>0$
and $\kappa_i=0$ for a slack payment inequality.

\subsection{Exact Path Pricing}
\label{subsec:gactg_shortest_path}

Let $c_{jl}$ be the contribution of arc $l$ to the private cost
$c_{jh}$ in \eqref{eq:job_private_cost}. Additivity of costs and resource
profiles gives, for $j\in\mathcal J_i$,
\begin{subequations}\label{eq:gactg_path_pricing}
\begin{align}
c_{jh}&=\sum_{l\in h}c_{jl},~ h\in\mathcal{H}_j,
 \label{eq:gactg_arc_additivity}\\
w_{jl}&=\kappa_ic_{jl}
 +\sum_{t\in\mathcal{T}}\big[\alpha_{i,t}G_{jl,t}+\beta_{i,t}B_{jl,t}
 \nonumber\\*
&~+(\psi_{i,t}+\Delta\lambda_{n(i),t})\eta_ip_{jl,t}^{\mathrm{IT}}\big],
 ~ l\in\mathcal{L}_j, \label{eq:gactg_arc_weight}\\
\bar c_{jh}&=\sum_{l\in h}w_{jl}
 -\kappa_ic_{j,h_j^{\mathrm{base}}}-\sigma_j,~ h\in\mathcal{H}_j.
 \label{eq:gactg_baseline_cost}
\end{align}
\end{subequations}
The last two terms in \eqref{eq:gactg_baseline_cost} are job constants.
Set the source distance to zero and others to infinity. Relax arcs in
topological order to find the minimum reduced-cost terminal path in
$O(|\mathcal S_j|+|\mathcal L_j|)$ per job on preordered lists, even with
negative weights. Each round adds negative reduced-cost paths and
resolves the RMP. In the RTS graphs, reachability and completion-time
bounds prune states outside feasible paths (Appendix~\ref{app:repro}).

\subsection{Bounds and Integer Recovery}
\label{subsec:gactg_integer}
\label{subsec:gactg_bounds}

Let $\mathcal D$ collect rounds with an optimal RMP and exact
full-family pricing at its current unsmoothed dual solution,
before path fixing. In round $d$, the RMP objective is
$Z_{\mathrm{RMP}}^{(d)}$. The bound is
\begin{subequations}\label{eq:gactg_bound_group}
\begin{align}
\bar c_j^{\min,(d)}
&=\min_{h\in\mathcal{H}_j}\bar c_{jh}^{(d)},~ j\in\mathcal{J},
 \label{eq:gactg_minimum_reduced_cost}\\
LB^{(d)}
&=Z_{\mathrm{RMP}}^{(d)}
 +\sum_{j\in\mathcal{J}}\min\{0,\bar c_j^{\min,(d)}\},
 \label{eq:gactg_lower_bound}\\
LB^{\mathrm{best}}&=\max_{d\in\mathcal D}LB^{(d)}.
 \label{eq:gactg_best_bound}
\end{align}
\end{subequations}
\begin{proof}[Validity of \eqref{eq:gactg_lower_bound}]
Set $\delta_j=\min\{0,\bar c_j^{\min,(d)}\}$ and replace $\sigma_j$ with
$\sigma_j+\delta_j$. Every full-master path then has dual slack
$\bar c_{jh}^{(d)}-\delta_j\ge0$. The other dual constraints still hold.
Strong duality gives $Z_{\mathrm{RMP}}^{(d)}$ for the original dual objective.
Each convexity row has right-hand side one, so the shift adds
$\sum_j\delta_j$ to that objective. Weak duality bounds the full LP
objective from below, which in turn bounds the integer optimum.
Taking the maximum preserves validity. When all minima are nonnegative,
the original dual and the RMP solution certify the full LP optimum.
\end{proof}

Several dual values may be optimal at $\Delta C_i=0$. This can slow
column generation. We also use Wentges smoothing to find
columns~\cite{wentges1997weighted}.
\begin{equation}
\widetilde\mu^{(d)}
=\vartheta\mu^{\mathrm c}+(1-\vartheta)\mu^{(d)},
~ 0\le\vartheta<1.
\label{eq:gactg_smoothed_duals}
\end{equation}
The current duals are $\mu^{(d)}$, and $\mu^{\mathrm c}$ starts at
$\mu^{(1)}$ and updates when $LB^{\mathrm{best}}$ improves.
Smoothed and dive duals generate candidate columns. Heuristic and
smoothed pricing iterations do not contribute to the global certificate.
The bounds remain valid at early termination.

Dive-and-price fixes integral LP selections, then repeatedly fixes a job
to the path with the largest fractional weight and resumes pricing. An infeasible
fix is reversed and its column excluded. A restricted MILP searches all
generated columns with dive restrictions removed. Replay and redispatch
verify the best integer plan, including the baseline candidate, and give
cost $UB$ with absolute gap $UB-LB^{\mathrm{best}}$.

\section{Numerical Evaluation}
\label{sec:evaluation}

\subsection{Test System and Workloads}
\label{subsec:experimental_setup}

The Reliability Test System of the Grid Modernization Laboratory
Consortium (RTS-GMLC) has 73 buses, 120 branches, and 73 thermal units
totaling 8.1~GW~\cite{barrows2020rts}. We use continuous line ratings,
published ramps and minimum up/down times, and convex piecewise-linear
heat-rate costs. Photovoltaics (PV), wind, concentrating solar, and
hydro can be spilled at zero penalty. Grid cost covers fuel and
no-load costs.
Before scheduling, we select dates from the hourly day-ahead series by
peak load (Aug.~26), highest PV share of demand (Mar.~29), highest
wind energy (Jan.~7), and near-median daily energy (Dec.~8). Reference cases use
15-min real-time averages ($\Delta=0.25$~h, $T=96$) and commitment
based on those inputs and campus baselines.

Three AI campuses host 15, 25, and 35 jobs at 230-kV buses 121, 215,
and 323, which have the largest sum of incident ratings in their areas.
Their nominal baseline peaks (150, 300, and 500~MW) total 10.85\%
of the Aug.~26 15-min system peak including campuses. We sample
gang sizes and durations from Acme pretraining jobs and scale gangs
to match campus peaks~\cite{hu2024acmetrace}. The 10th and 90th
percentiles across 750 jobs are 2,450.91 and 29,410.90 GPU equivalents. These are
power-scaled computing capacities rather than observed physical GPU
counts for individual jobs. The resulting profiles assume proportional
scaling of computing capacity, power, and checkpoint state size.
Campus-scale operation profiles are scenario inputs.

Durations are clipped to 2--16~h. Training credits one work unit per
slot, with quotas shortened to fit arrival, baseline operations,
and deadlines. Initially active jobs (probability 0.4) have fresh
checkpoints. Other jobs arrive in the first half of the day.
H100 idle and Llama-2 70B low-rank adaptation (LoRA) measurements
inform GPU power~\cite{vercellino2026genai}. Table~\ref{tab:params}
adds assumed non-GPU power, operation and storage profiles, and
overhead.

The baseline starts new jobs at arrival. Jobs train continuously between
writes, save after two hours of unprotected work, and finalize.
We add 30\% to baseline durations from arrival
and round up to whole slots for deadlines. GPU capacity is 10\% above
peak baseline occupancy, and interconnection limits are 125\% of
nominal baseline peaks. The two-hour work cap is 8 units at 15~min
or 24 at 5~min. Idle facility power is 1.2 times idle IT power.
Each RTS-GMLC job has one available execution mode. Appendix~\ref{app:small} checks
recovery and mode availability.

\begin{table}[t]
\centering
\caption{Declared execution and contract parameters}
\label{tab:params}
\begin{tabular}{ll}
\hline
Idle power / training increment & 0.25 / 0.86 kW/GPU \\
Hold / write / restore increment & 0.08 / 0.12 / 0.20 kW/GPU \\
Write duration (nominal / reserved) & 5 / 15 min \\
Restore / initialization duration & 15 min$^\dagger$ \\
Checkpoint shard per GPU & 0.14 GB ($10^7$ parameters) \\
Storage bandwidth per campus & 200 GB/s \\
Unprotected work $\overline U_j$ & 8 units (2 training h) \\
Marginal overhead factor $\eta_i$ & 1.2 \\
Contract tariff $\rho_{i,t}$ & \$70/MWh \\
GPU-time charge & \$0/GPU-h (reference) \\
\hline
\multicolumn{2}{l}{\footnotesize $^\dagger$Above 8192 GPU equivalents, restore and initialization take 30 min.}
\end{tabular}
\end{table}

Operation power and checkpoint size scale with gang size. Checkpoints
use 14 bytes per parameter in decimal GB. Energy and I/O use the
reserved duration in Table~\ref{tab:params}. Finalization uses the
write profile, and initialization uses restore power with zero I/O.
Shutdown takes zero slots. A checkpoint of $S_j$ GB uses
$S_j/(3600\Delta\tau)$~GB/s over its $\tau$ reserved slots.

\subsection{Comparison Policies}
\label{subsec:comparison_methods}

Joint chooses each write within the unsaved-work limit. The first-write
variant chooses the initial work cap and write point for each job.
Later writes use a common two-hour training interval. Both select paths,
dispatch, and payments jointly.

The controls share jobs, baselines, campus limits, and payment rules.
We fix each policy's job paths and redispatch generation under the
original line limits and commitment (Appendix~\ref{app:baselines}).
\begin{itemize}[leftmargin=*,nosep]
\item \textbf{Base} follows the agreed fixed-cadence schedule.
\item \textbf{Periodic} requires two hours of unsaved work before each intermediate
write, including the first~\cite{mohan2021checkfreq,lu2026battery}.
\item \textbf{HoldOnly} uses Periodic's cadence and keeps GPUs assigned until completion.
\item \textbf{CompObl} selects paths to minimize grid operating cost, then settles payments.
\item \textbf{NetObl} relaxes line limits during path selection.
\item \textbf{Price} minimizes the energy cost of the selected paths at fixed
baseline locational marginal prices plus payments~\cite{tsg2021incentiveIDC}.
\item \textbf{Checkpoint-state scheduling (CS)} adapts off-GPU waiting at preset
progress boundaries~\cite{tong2026execution} to the single-mode jobs.
\item \textbf{Envelope} chooses a campus power target subject to dispatch and
cumulative energy constraints, then fits job paths to that target~\cite{tsg2021aggregatedIDC}.
\end{itemize}

\subsection{Grid Limits Shape Feasible Schedules}
\label{subsec:value}
A pause helps the grid only if it can serve the load when jobs resume.
In peak sample~7, redispatch of NetObl requires
a 6.42-MW increase if only circuit 107--203's rating can change
(Appendix~\ref{app:small}). Joint, first-write, Periodic,
and HoldOnly serve the same jobs at the original ratings
and commitment.
Median Joint savings are larger on the solar and wind days
(Table~\ref{tab:full_comparison}).

\begin{table}[!ht]
\centering
\caption{Redispatch feasibility and net savings by policy}
\label{tab:seed_coverage}\label{tab:full_comparison}
\footnotesize
\setlength{\tabcolsep}{2pt}
\begin{tabular}{@{}lrrrrr@{}}
\hline
 & & \multicolumn{4}{c}{Median share of Joint net savings (\%)} \\
Policy & Feasible (/40) & Solar & Wind & Median energy & Peak\\
\hline
Base & 40 & 0.00 & 0.00 & 0.00 & 0.00 \\
Periodic & 40 & 94.31 & 92.10 & 85.38 & 79.14 \\
HoldOnly & 40 & 74.41 & 70.23 & 62.64 & 22.61 \\
CompObl & 40 & 95.41 & 88.93 & 47.04 & 1.25 \\
NetObl & 39 & 98.97 & 85.06 & 99.33 & 63.68 \\
Price & 32 & 71.70 & 52.71 & -44.41 & 23.87 \\
Envelope & 32 & -19.76 & -28.17 & -631.85 & -939.38 \\
CS & 40 & 94.15 & 90.35 & 83.12 & 79.14 \\
\cline{1-6}
 & & \multicolumn{4}{c}{Median net savings (\$k)} \\
\textbf{Joint (proposed)} & 40 & 13.12 & 11.89 & 2.40 & 1.20 \\
\hline
\end{tabular}
\par\vspace{1pt}
\begin{minipage}{\columnwidth}\footnotesize
The feasible count uses fixed-load redispatch. For each control, we
divide its net savings by Joint's in each feasible case, then take the
median for that day. Negative shares indicate costs above Base.
\end{minipage}
\end{table}

\begin{figure}[!b]
\centering
\subfloat[Demand (solar day).\label{fig:e1_demand}]{%
\includegraphics[width=0.495\columnwidth]{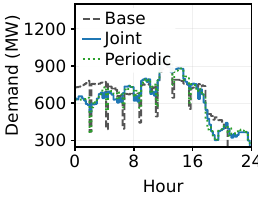}}\hspace*{0.15em}%
\subfloat[Mean campus-bus LMP.\label{fig:e1_lmp}]{%
\includegraphics[width=0.495\columnwidth]{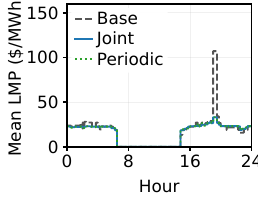}}
\caption{Campus demand and mean campus-bus prices on the solar reference day.
Joint avoids 442.23~MWh of curtailment relative to Base.}
\label{fig:e1}
\end{figure}

On the solar day, release and recovery
move training toward lower prices and surplus renewable output
(Fig.~\ref{fig:e1}).

\label{subsec:envelope}
Price can concentrate demand in low-price slots where capacity is limited.
Envelope fits job paths to its dispatch-feasible target, which can change
the load in constrained slots. Both controls fail some fixed-load
replays. Envelope's median savings share is negative on every reference day (Table~\ref{tab:full_comparison}).

The peak day has the smallest median Joint savings and the lowest
median share for feasible NetObl plans. Transfer limits therefore affect
the cost of recovery even when redispatch can serve the scheduled load.

\subsection{Scheduling and Release Drive Most Savings}
\label{subsec:release_value}
Holding shifts training while keeping GPUs assigned. Release lets other
jobs use those GPUs, and a checkpoint preserves work for recovery.
Median retained savings rise from HoldOnly to Periodic
in all four day types (Table~\ref{tab:seed_coverage}).
In solar sample~1, holding and release account for 92.38\% of the
Joint plan's net savings (Appendix~\ref{app:timing}).
In the three-bus ablation, restoration lets jobs release their GPUs
and later resume from saved work. The gain is clearest when the
corridor binds (Appendix~\ref{app:small}).

\subsection{Checkpoint Timing and Operating Sensitivities}
\label{subsec:checkpoint_timing}

\noindent \textbf{\textit{Choosing the first write.}}
An earlier first write can preserve progress before a planned release.
With a common two-hour training interval after the first write,
first-write retains at least 90.37\% of optimal Joint net
savings in each of the 40 reference cases (Appendix~\ref{app:timing}).
Table~\ref{tab:first_write} gives the share of Joint plan savings retained by
first-write for sample~1 on each reference day.

\begin{table}[!ht]
\centering
\caption{Net savings retained by first-write for sample 1}
\label{tab:first_write}
\footnotesize
\setlength{\tabcolsep}{3pt}
\begin{tabular*}{\columnwidth}{@{\extracolsep{\fill}}lrrrr@{}}
\hline
Reference day & Solar & Wind & Median energy & Peak \\
\hline
Retained (\%) & 98.28 & 98.67 & 96.95 & 97.48 \\
\hline
\end{tabular*}
\end{table}

In the peak-line case, one job first writes in the same slot under
both policies. Joint releases it with less saved work and restores before the line binds.
First-write restores during a binding slot (Appendix~\ref{app:timing}).

\begin{figure}[!ht]
\centering
\subfloat[First job's progress.\label{fig:timing_progress}]{%
\includegraphics[width=0.495\columnwidth]{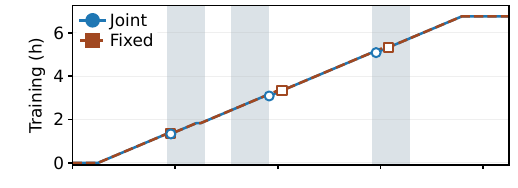}}\hspace*{0.15em}%
\subfloat[Campus demand.\label{fig:timing_power}]{%
\includegraphics[width=0.495\columnwidth]{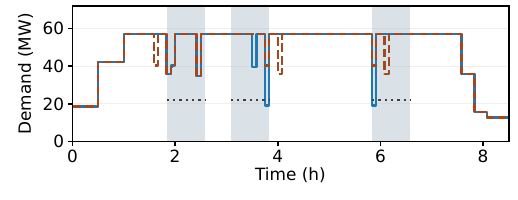}}
\caption{Equal write counts for two jobs with 5-min slack. Shading marks
three uneven import windows. Markers show completed writes and dotted
lines show available imports.}
\label{fig:timing_windows}
\end{figure}

\noindent \textbf{\textit{Choosing each write.}}
Fig.~\ref{fig:timing_windows} compares Joint with a tuned fixed policy
at equal write counts in a constructed three-bus test. The tuned
policy chooses each job's first write and repeat interval.

The final segment has a two-hour cap on unsaved work. Under both
policies, jobs retain their GPUs and can wait and write synchronously.

With little time to wait, Joint places job~1's second write and
job~2's first write in separate import windows. In the two-job case,
these low-power slots avoid 2.70~MWh of local
generation at equal facility energy and compensation.

The tuned fixed cadence aligns writes in uniform windows. At 30\%
deadline slack, both policies in these tests can wait to avoid local
generation (Appendix~\ref{app:timing}).

\noindent \textbf{\textit{Write-power sensitivity.}} We average write and training
power over each intermediate checkpoint reservation using the blocking
fraction, with fixed work credit and durability rules.
In the case in Fig.~\ref{fig:timing_windows}, choosing each write saves
\$539.47 with a full-slot dip and \$1.25 with 0.59~s of blocking
(Appendix~\ref{app:timing}).

\subsection{Costs and Resource Limits Affect Savings}
\label{subsec:conditions}
A schedule that cuts generation costs can require enough payment
to erode its net savings.
At \$0.25/GPU-h in solar sample~1, Joint and Periodic
retain GPUs until completion with equal daily energy. First-write
captures most of the timing benefit (Appendix~\ref{app:timing}).
Longer-than-planned restores cause missed deadlines in the recovery
duration check (Appendix~\ref{app:small}).

\noindent \textbf{\textit{Campus location and size.}}
Repartitioning the same 75 jobs with fixed total capacity changes which
jobs share GPUs and storage and which buses serve their recovery load.
Joint savings fall as solar-day sites are added. Wind-day savings
reach a minimum at nine sites and partly recover at twelve
(Appendix~\ref{app:timing}, \textit{Campus siting}).

\noindent \textbf{\textit{Magnitude of savings.}}
A separate 366-day realized-input test for 2020 (sample~1) yields
\$1.55M (0.27\%) in net savings and 47.37~GWh less curtailment.
Work and checkpoints reset daily. Joint and Base share
each day's fixed unit commitment. The percentage uses baseline system
dispatch cost excluding startup charges.
Participating campuses supply the flexibility. This ratio does not directly measure the economic value of each
delivered MW of flexibility. Read it alongside avoided curtailment,
network feasibility, and compensation costs.

\subsection{Savings Depend on Forecasts and Commitment}
\label{subsec:workload_calendar}
\label{subsec:forecast}
Day-ahead (DA) plans repeat hourly forecasts across four slots, and
conditional perfect foresight (PF) uses realized profiles. In the first
panel of Table~\ref{tab:forecast_policies}, PF chooses paths under
realized Base commitment with DA slow units. Replay fixes paths
and slow units and recommits fast combustion turbines (CTs). The second
panel uses a common retrospective commitment that serves both baselines.
Costs include fuel, no-load, startup, and payments.

\begin{table}[!ht]
\centering
\caption{Startup-inclusive realized savings (\$k)}
\label{tab:forecast_policies}
\footnotesize
\setlength{\tabcolsep}{7pt}
\begin{tabular}{lrrrr}
\hline
 & \multicolumn{2}{c}{Solar} & \multicolumn{2}{c}{Peak} \\
Policy & DA & PF & DA & PF \\
\hline
\multicolumn{5}{l}{\textit{Recommitted fast units}} \\
Joint & 10.24 & 55.40 & -16.57 & 0.70 \\
Periodic & 13.77 & 52.52 & -14.97 & 0.26 \\
Price & 9.12 & 52.12 & -48.97 & 0.02 \\
CS & 14.04 & 50.22 & -14.90 & 0.12 \\
\multicolumn{5}{l}{\textit{Common commitment}} \\
Joint & -- & 37.73 & -- & 0.66 \\
First-write & -- & 37.40 & -- & 0.65 \\
\hline
\end{tabular}
\par\vspace{1pt}
\begin{minipage}{\columnwidth}\footnotesize
Each panel uses its own realized Base. Dashes mark infeasible
redispatch. Startup charges count off-to-on transitions after slot~0. The
recommitted panel allows different initial fast-unit states.
\end{minipage}
\end{table}

In the recommitted peak replay, DA Joint costs more than
Base and incurs higher startup charges. On the solar day,
CS and Periodic lead under DA, and Joint
leads under PF. With common commitment, DA Joint and
first-write exceed available supply on both days even after
relaxing line and ramp limits.

\noindent \textbf{\textit{Rolling replay.}}
In a separate four-day replay, policies share a fixed daily commitment serving
forecast Base at 5\% higher background load. Rolling planners blend
forecasts with recent observations and replan every four hours. Both
DA planners retain Base within budget. The median day alone is
feasible for all policies. Relative to their DA counterparts, rolling
first-write and Joint cut wind-day unserved energy by
50.41 and 49.44~MWh and deepen a peak shortage that
persists at campus idle load.

\subsection{Scalability and Solution Quality}
\label{subsec:computation}
Table~\ref{tab:e6} compares GACTG and arc-flow across job counts and
slot lengths on the solar day. Times cover model construction and solving.

\begin{table}[t]
\centering
\caption{Solar-day computation by workload size and slot duration}
\label{tab:e6}
\footnotesize
\setlength{\tabcolsep}{2pt}
\begin{tabular}{rrrrrrr}
\hline
 & & & \multicolumn{2}{c}{GACTG} & \multicolumn{2}{c}{Arc-flow} \\
Jobs & Slot (min) & Arcs (M) & Time (s) & Gap (\%) & Time (s) & Gap (\%) \\
\hline
24 & 15 & 0.40 & 48.02 & 0.14 & 1815.78 (TL) & 0.06 \\
48 & 15 & 0.77 & 143.33 & $<0.01$ & 727.16 & $<0.01$ \\
75 & 15 & 1.39 & 20.49 & $<0.01$ & 564.80 & $<0.01$ \\
150 & 15 & 2.78 & 54.29 & $<0.01$ & 897.07 & $<0.01$ \\
300 & 15 & 5.89 & 35.13 & $<0.01$ & 323.93 (MEM) & -- \\
12 & 5 & 6.33 & 310.70 & 0.04 & 284.92 (MEM) & -- \\
24 & 5 & 9.10 & 248.08 & 0.01 & 404.53 (MEM) & -- \\
48 & 5 & 17.23 & 228.88 & $<0.01$ & 400.28 (MEM) & -- \\
\hline
\end{tabular}
\par\vspace{1pt}
\begin{minipage}{\columnwidth}\footnotesize
GACTG uses one thread (30~min for pricing and diving, 120~s for the
restricted MILP). Arc-flow uses eight threads (30-min solve limit, 22~GB
per run). Times exclude common preprocessing (2.40--12.69~min).
Gaps are relative to feasible cost. TL and MEM mark time and memory stops.
\end{minipage}
\end{table}

\begin{figure}[!ht]
\centering
\subfloat[Full bound history.\label{fig:e6_conv}]{%
\includegraphics[width=0.495\columnwidth]{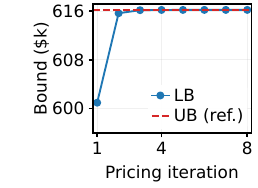}}\hspace*{0.15em}%
\subfloat[Zoom, iterations 4--8.\label{fig:e6_zoom}]{%
\includegraphics[width=0.495\columnwidth]{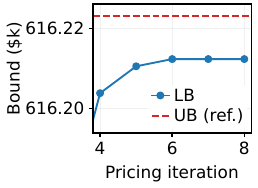}}
\caption{Global lower bounds for 150 jobs, with a zoom and the final feasible upper-bound reference.}
\label{fig:e6}
\end{figure}

GACTG finds feasible schedules throughout. Arc-flow hits memory limits
with finer slots or more jobs. The 150-job bound improves mainly in
iteration~2 (Fig.~\ref{fig:e6}).

\noindent \textbf{\textit{Matched-budget results.}}
With one thread, a nominal 90-s budget, and a 3-GB soft memory limit,
the tests start from Base.
For 75 jobs, GACTG saves \$10,722.93 in 21.88~s (gap \$9.38).
For 150 jobs, GACTG saves \$12,914.24 in 12.71~s (gap \$10.84).
Arc-flow retains Base, stopping at the soft memory limit (80.57~s) and time limit (95.40~s), respectively.

\noindent \textbf{\textit{118-bus network.}}
We use line ratings and generation costs from the Power Grid Library
(PGLib)~\cite{pglib2023}. Net savings increase with solar availability
in the matched 75-job cases (Appendix~\ref{app:network118}).

\section{Conclusion}
\label{sec:conclusion}
With realized inputs, joint scheduling serves the peak workload under
the original ratings. A flexible first write and a common later training
interval retain most net savings in the reference cases. Per-write
control reduces local generation in cases with tight deadlines and
uneven import windows. Forecast errors can undo savings on the peak day. Future work will measure write power, time to durable save, and restore time with shared campus storage. Broader forecast replays with adjustable
unit commitment can test when rolling coordination reduces shortfalls and costs.

\appendix
\subsection{Graph and Reproducibility Details}
\label{app:repro}

\noindent \textbf{\textit{Arc-flow equivalence.}} Join completion states to one sink
with zero-cost, zero-use arcs. Balance and acyclicity allow a
positive-flow source-to-sink path. Subtract its minimum flow and
repeat. Each step eliminates an arc, and acyclicity precludes
circulation. Path weights thus sum to one, and integer flow selects
one path. Conversely, convex combinations give unit flows with the
same additive profiles and costs.

\noindent \textbf{\textit{Exact graph reductions.}} Contract zero-use pre-arrival waits
and keep states on source-to-completion paths. In the single-mode RTS
graphs, an active state requires at least $W_j-x$ training slots plus
finalization. New jobs also need initialization. For off jobs, take
the faster completion bound of initialization from zero and compatible
recovery from $q$. Done states need zero time. Discard a state if its
bound exceeds the time to deadline.

GACTG uses $\vartheta=0.5$. Up to two RTS arc-flow runs execute
concurrently. Arc-flow uses one thread for the 118-bus tests.

\noindent \textbf{\textit{Timing protocol.}} For the matched-budget tests, we time construction and solving after
common preprocessing and GACTG pricing warm-up. Replay and redispatch
are untimed.

\subsection{Mechanism and Execution Checks}
\label{app:small}

\noindent \textbf{\textit{Three-bus network.}} We cross 24 four-job workloads with two load
windows and two corridor limits on four logical GPUs. Jobs start
without checkpoints. Forbidding intermediate writes and allowing them
without checkpoint restoration (WriteOnly) yield the same optima. Restoration adds value mainly
when the corridor binds (Table~\ref{tab:write_only_boundaries}).
In one case, zero-power holding needs six GPUs to keep the training
times of write--release--restore. Faster modes change the cost of
recomputation and the value of recovery.

\begin{table}[!ht]
\centering
\caption{Recovery-enabled increments across 96 three-bus scenarios}
\label{tab:write_only_boundaries}
\begin{tabular}{lrrr}
\hline
Condition & Positive & Zero & Unresolved \\
\hline
Reference & 47 & 48 & 1 \\
Zero incremental holding power & 36 & 60 & 0 \\
\hline
\end{tabular}
\end{table}

\noindent \textbf{\textit{IEEE 30-bus network.}} Across 216 cases with three site/size
settings and three ramp limits, quadratic-cost certificates show higher
costs than Joint in 111 NetObl cases (all at bus~8) and
53 CompObl cases, even with tie breaking favorable to each control.

\noindent \textbf{\textit{RTS-GMLC dispatch.}} Removing line limits makes seven of 17
failed replays feasible. Five exceed generation capacity even after
relaxing line and ramp limits.
For peak NetObl, circuit 107--203 needs
6.42~MW more when other ratings are fixed. With all ratings adjustable,
the minimum maximum overload is 2.82~MW.

\noindent \textbf{\textit{Single-GPU execution.}} Nine RTX~3070 runs pass state and loss
checks (43M parameters, three seeds, 256-step packets). Recovery avoids
one recomputed packet and finishes 14.43--27.24~s earlier than reset.

\noindent \textbf{\textit{Dual-GPU recovery.}} Two jobs share an A100 node's network volume.
All 18 restores match uninterrupted training at the next update and
completion. Concurrent restores have longer median per-job readiness
times than isolated restores. From the common request, the pair completes
sooner with concurrent restores than with staggered starts in all three trials.

\noindent \textbf{\textit{Recovery slack.}} Adding 15~min to each of 25 restores
and shifting later operations makes 20 jobs late on the 5-min solar grid.
Planning the longer restores in prior off-GPU waits meets deadlines
and network limits for \$1,516.36 more.

\subsection{Comparison Models}
\label{app:baselines}

\noindent \textbf{\textit{Path restrictions.}} Joint uses $\mathcal H_j$.
Periodic permits intermediate writes only at $x-q=\overline U_j$.
First-write selects one of eight caps by exact pricing on the
15-min grid. At zero saved progress, including after reset, this cap
limits training and sets the write point. Later writes use the common
two-hour cap. With the same jobs, line limits, profiles, and payments,
$C_{\mathrm{Joint}}^\star\le C_{\mathrm{first\text{-}write}}^\star
\le C_{\mathrm{Periodic}}^\star$.
CS permits off-GPU waits at preset boundaries and forbids holding,
abrupt release, and reset.

\noindent \textbf{\textit{Objective and network controls.}} CompObl minimizes
$\mathcal C_{\mathrm{grid}}$ over the full feasible set, then settles
payments. NetObl solves $\mathcal P$ without
\eqref{eq:line_capacity}, then restores line limits for evaluation.
For frozen baseline LMPs $\lambda_{n(i),t}^{\mathrm{base}}$,
Price solves
\begin{equation}
\min~\Delta\sum_{i,t}\lambda_{n(i),t}^{\mathrm{base}}
\eta_i\sum_{j\in\mathcal J_i,h\in\mathcal H_j}p_{jh,t}^{\mathrm{IT}}z_{jh}
+\sum_i M_i,
\label{eq:price_control}
\end{equation}
subject to trajectory selection, GPU, I/O, interconnection, and payment
constraints. Its selection stage has no dispatch or line-flow variables.

\noindent \textbf{\textit{Aggregate-energy control.}} Envelope selects incremental
IT power $v_{i,t}$ jointly with dispatch, minimizing
$\mathcal C_{\mathrm{grid}}$. Let $E_{i,t}^{\mathrm{early}}$ and
$E_{i,t}^{\mathrm{late}}$ be cumulative energies from baseline operation
sequences placed at their earliest starts and latest completions.
Along with dispatch and interconnection limits, the envelope imposes
\begingroup\setlength{\jot}{1pt}
\begin{subequations}\label{eq:envelope_control}
\begin{align}
&0\le v_{i,t}\le\overline v_i,~
P_{i,t}^{\mathrm{DC}}=P_{i,t}^{\mathrm{idle}}+\eta_i v_{i,t},~\forall i,t,\\
&E_{i,t}^{\mathrm{late}}\le\Delta\sum_{t'=0}^{t}v_{i,t'}
\le E_{i,t}^{\mathrm{early}},~\forall i,t,\\
&\Delta\sum_t v_{i,t}=E_i^{\mathrm{base}},~\forall i.
\end{align}
\end{subequations}
\endgroup
Here $\overline v_i$ is all-GPU training power and
$E_i^{\mathrm{base}}$ is baseline daily incremental IT energy.
A second optimization selects one path per job under GPU and I/O limits
to minimize $\Delta\sum_{i,t}|\eta_i\sum_{j\in\mathcal J_i,h\in\mathcal H_j}
p_{jh,t}^{\mathrm{IT}}z_{jh}-\eta_i v_{i,t}|$.
The replayed paths determine the reported cost and dispatch feasibility.

\subsection{Additional Timing and Operating Tests}
\label{app:timing}

\noindent \textbf{\textit{Solar increments.}}
In solar sample~1, net savings rise by \$7,476.30 from Base
to HoldOnly and by \$2,430.55 from HoldOnly to
Periodic. First-write and later-write control add
\$632.99 and \$184.38, respectively.

\noindent \textbf{\textit{Two-hour limit.}} In each of the 40 reference cases,
choosing every write saves at most \$433.86 over first-write
and at most 9.63\% of optimal Joint net savings.
On 12 days sampled by quarter and baseline congestion, first-write retains
90.61--99.17\% of Joint plan net savings.
Table~\ref{tab:first_write_costs} gives cost differences for Table~\ref{tab:first_write}'s
four cases and two additional peak cases.

\begin{table}[!ht]
\renewcommand{\arraystretch}{0.95}
\centering
\caption{Cost of fixing later writes at the common interval (\$)}
\label{tab:first_write_costs}
\footnotesize
\setlength{\tabcolsep}{3pt}
\begin{tabular}{lrr}
\hline
Case & Plan difference & Optimal difference \\
\hline
Solar & 184.38 & [174.33, 192.48] \\
Wind & 151.90 & [140.46, 166.19] \\
Median & 85.89 & [79.08, 94.20] \\
Peak & 28.60 & [19.78, 35.49] \\
Peak sample 2 & 89.70 & [87.97, 91.02] \\
Peak sample 7 & 44.47 & [36.89, 52.84] \\
\hline
\end{tabular}
\par\vspace{1pt}
\begin{minipage}{\columnwidth}\footnotesize
Differences are first-write minus Joint. The first four cases share a
workload. For a control P versus Joint (J), brackets use
$[LB_{\mathrm P}-UB_{\mathrm J},\,UB_{\mathrm P}-LB_{\mathrm J}]$
to bound $C_{\mathrm P}^\star-C_{\mathrm J}^\star$.
Peak samples~2 and~7 retain 90.50\% and 95.81\%
of Joint plan savings, respectively.
\end{minipage}
\end{table}

\noindent \textbf{\textit{Tuned cadence.}} Exact pricing tests a first and a repeat
interval per job. The final segment uses the common two-hour cap.
Choosing every write cuts optimal cost by \$54.20--61.20 in peak sample~2
and \$34.53--52.84 in the peak-line case.

\noindent \textbf{\textit{Peak-line recovery.}} The bus-215 job first writes in slot~30 in
both plans. Joint releases at 32 saved units, restores at~75, and writes
at~83. First-write releases at 40 units and restores at~84. Line
107--203 binds at 175~MW in slots~83--84 in both plans.
Across the full schedules, Joint cuts generation cost by \$119.27
and raises payments by \$74.80 relative to first-write.

\noindent \textbf{\textit{Slot length and operation profiles.}} Repeating 15-min inputs
on the 5-min grid preserves energy and adds decisions and ramp checks. Work,
deadlines, checkpoint sizes, capacities, and commitment are common.

\noindent \textbf{\textit{Equal write counts.}} The three-bus cases cross one or two jobs,
three uniform or uneven 45-min windows, and 5-min or 30\% slack.
Jobs keep GPUs, make three 5-min writes, and finalize in 15~min.
All four segments have a two-hour cap. The tuned fixed policy requires
equal training work in the two interior segments (Table~\ref{tab:equal_writes}).

\begin{table}[!ht]
\renewcommand{\arraystretch}{0.95}
\centering
\caption{Reductions from choosing each write at equal write counts}
\label{tab:equal_writes}
\footnotesize
\setlength{\tabcolsep}{4pt}
\begin{tabular}{llrrrr}
\hline
& & \multicolumn{2}{c}{Uniform windows} & \multicolumn{2}{c}{Uneven windows} \\
Jobs & Slack & Cost (\$) & MWh & Cost (\$) & MWh \\
\hline
1 & 5 min & 0.00 & 0.00 & 166.13 & 0.83 \\
2 & 5 min & 0.00 & 0.00 & 539.47 & 2.70 \\
1 & 30\% & 18.24 & 0.00 & 0.00 & 0.00 \\
2 & 30\% & 0.00 & 0.00 & 0.00 & 0.00 \\
\hline
\end{tabular}
\par\vspace{1pt}
\begin{minipage}{\columnwidth}\footnotesize
MWh is local generation avoided relative to the tuned fixed policy.
Slack is relative to the agreed schedule. The \$18.24 difference
reflects one extra hold under the fixed policy.
\end{minipage}
\end{table}

\noindent \textbf{\textit{Short write-power dips.}}
The two-job gain is \$315.65 at a 148.8-s blocking time~\cite{pasqualin2024async}.
At 0.59~s~\cite{wan2025bytecheckpoint}, both RTS baselines fail
redispatch under the original commitments. At shared commitments,
the gain from choosing every write over first-write falls from \$170.73--190.81
to at most \$11.07 (solar) and from \$37.27--52.39 to at most \$3.86 (peak-line).
Joint still saves \$10,217.65 and \$1,166.60 on the two days.
The peak-line plan meets its binding 175-MW rating.

\noindent \textbf{\textit{Storage bursts.}}
Across 40 cases, GPU caps bound campus 15-min average I/O by 67.31~GB/s.
Applying measured save and restore times~\cite{wan2025bytecheckpoint} to modeled
checkpoint sizes gives peak-line bursts of 810.95~GB/s (Base)
and 344.31~GB/s (both coordinated plans). Pacing these batches
to 200~GB/s takes 207.04~s and 87.90~s, respectively.
Joint's longest batch takes 129.49~s. All batches meet their reservations.

\begin{table}[t]
\renewcommand{\arraystretch}{0.95}
\centering
\caption{Fixed-work cost increment of Periodic over Joint (\$)}
\label{tab:matched_clock}\label{tab:profile_sensitivity}
\footnotesize
\setlength{\tabcolsep}{2pt}
\begin{tabular}{@{}rrrr@{}}
\hline
\shortstack{Slot/\\Write} & \shortstack{Restore/\\Power} &
\shortstack{Solar\\Plan [LB, UB]} & \shortstack{Wind\\Plan [LB, UB]} \\
\hline
15/15 & 1/1 & 817.37 [803.30, 825.47] & 762.39 [746.37, 776.68] \\
5/15 & 1/1 & 1008.02 [944.56, 1052.08] & 868.00 [833.60, 957.73] \\
5/10 & 1/1 & 499.56 [462.88, 597.00] & 1215.78 [1174.44, 1252.69] \\
5/5 & 1/1 & 421.55 [372.68, 474.30] & --- \\
5/5 & 0.5/1 & 416.82 [379.22, 460.03] & --- \\
5/5 & 2/1 & 312.45 [289.05, 355.97] & --- \\
5/5 & 1/0.5 & 575.00 [526.81, 631.36] & --- \\
5/5 & 1/1.5 & 307.08 [280.01, 346.01] & --- \\
\hline
\end{tabular}
\par\vspace{1pt}
\begin{minipage}{\columnwidth}\footnotesize
Slot and write durations are in minutes. Write includes finalization.
Restore duration and write, finalization, and restore power scale from
Table~\ref{tab:params}. Restore times round up to slots. Brackets bound
optimal differences. Dashes mark wind Base cases that fail redispatch
at the fixed commitment.
\end{minipage}
\end{table}

Joint costs less than
Periodic for feasible baselines (Table~\ref{tab:matched_clock}).
On the 5-min grid, raising write reservations from 10 to 15~min
raises the optimal gap on solar and lowers it on wind. The solar 5-
and 10-min bounds overlap. The five wind baselines with 5-min writes
resume training earlier and meet campus limits and deadlines. They
fail redispatch under the common commitment.

\noindent \textbf{\textit{Matched operating conditions.}} For sample~11 with fixed work,
a \$0.25/GPU-h fee cuts solar Joint releases from 24 to one.
At 15\% deadline slack, both days have lower total savings.
Across 19 matched solar and wind cases, GPU-time charges of \$0.05,
\$0.10, and \$0.25/GPU-h retain median shares of 68.57\%, 55.35\%, and
50.45\% of zero-charge savings.

In solar sample~1 at \$0.25/GPU-h, the Periodic and first-write
plans cost \$1,934.26 and \$385.05 more than the Joint plan, respectively.

\noindent \textbf{\textit{Release and strict cadence.}} An early write advances a bus-323
job's release and restore by 30~min on the solar reference day. Its
Periodic path adds \$45.34 in dispatch cost at equal compensation.
In the two-window case, fixed cadence and Joint tie at 5- and
15-min writes with the common four-hour cap on the final segment.

\noindent \textbf{\textit{Observed congestion windows.}} We use 15-min averages of the
negated published New York real-time zonal congestion-price
component~\cite{nyiso_realtime}.
Pairs of windows at \$50/MWh or more for 45~min or longer, with gaps
of 30~min to 6~h below threshold, occur on 19 New York City and 36 Long
Island days out of 725 complete days per zone. Of
72 RTS campus-days, one has a 45-min window at \$10/MWh and none has two.

\noindent \textbf{\textit{Campus siting.}} We assign the same 75 jobs to the highest-rated
connections at fixed total capacity, idle power, and commitment. Each
site serves its baseline peak (Table~\ref{tab:dc_count}).

\begin{table}[t]
\renewcommand{\arraystretch}{0.95}
\centering
\caption{Net savings with the same 75 jobs across campuses (\$k)}
\label{tab:dc_count}
\footnotesize
\setlength{\tabcolsep}{3pt}
\begin{tabular}{rrrrrrr}
\hline
 & \multicolumn{3}{c}{Solar} & \multicolumn{3}{c}{Wind} \\
Sites & Joint & Periodic & CS & Joint & Periodic & CS \\
\hline
3 & 10.72 & 9.91 & 9.87 & 11.41 & 10.65 & 10.49 \\
6 & 10.25 & 9.29 & 9.24 & 9.86 & 8.96 & 8.73 \\
9 & 10.01 & 9.16 & 9.09 & 8.69 & 7.74 & 7.53 \\
12 & 9.77 & 9.04 & 8.97 & 9.00 & 8.19 & 8.00 \\
\hline
\end{tabular}
\par\vspace{1pt}
\begin{minipage}{\columnwidth}\footnotesize
Gaps are below 0.10\%. After pricing converges, the restricted MILP uses
a $10^{-6}$ relative gap target and no time limit.
\end{minipage}
\end{table}

\subsection{Additional Network Results}
\label{app:network118}

The 118-bus tests use RTS daily shapes, all thermal units online,
and per-slot ramps of 10\% of output range. Campus peaks total 11\%
of nominal background load. Matched 75-job solar runs vary renewable
availability (Table~\ref{tab:network118}). Arc-flow reaches the time limit
with Base as its incumbent in all five solar runs, with a \$18.68 gap
for 24 jobs and zero lower bounds otherwise.

\begin{table}[t]
\renewcommand{\arraystretch}{0.95}
\centering
\caption{Constructed 118-bus savings and absolute certificate gaps}
\label{tab:network118}
\footnotesize
\setlength{\tabcolsep}{2pt}
\begin{tabular}{rlrrrrrr}
\hline
 & & Renewable & Arcs & \multicolumn{2}{c}{GACTG} & \multicolumn{2}{c}{Time (s)} \\
Jobs & Day & peak (\%) & (M) & Saving (\$) & Gap (\$) & GACTG & Arc-flow \\
\hline
24 & Solar & 30 & 0.52 & 17.27 & 1.43 & 8.74 & 305.80$^\dagger$ \\
75 & Solar & 30 & 1.26 & 26.06 & 1.61 & 10.81 & 310.99$^\dagger$ \\
150 & Solar & 30 & 2.52 & 47.14 & 1.01 & 36.51 & 324.56$^\dagger$ \\
75 & Wind & 30 & 1.26 & 25.08 & 1.06 & 15.11 & -- \\
75 & Solar & 60 & 1.26 & 304.37 & 0.96 & 9.28 & 311.76$^\dagger$ \\
75 & Solar & 90 & 1.26 & 1054.84 & 9.35 & 14.40 & 311.04$^\dagger$ \\
\hline
\end{tabular}
\par\vspace{1pt}
\begin{minipage}{\columnwidth}\footnotesize
Renewable peaks use nominal background load. Both methods use one thread.
Arc-flow limits are 300~s and 12~GB. $^\dagger$ indicates a time limit.
A dash marks a skipped run. Times include construction and solving after
common preprocessing. Periodic and CS retain 47.42--87.72\% of Joint savings.
\end{minipage}
\end{table}

\section*{Acknowledgment}
The authors used large language models (LLMs) to assist with manuscript polishing and grammar checking.

\vfill


\begin{thebibliography}{99}
\providecommand{\url}[1]{#1}
\csname url@samestyle\endcsname
\providecommand{\newblock}{\relax}
\providecommand{\bibinfo}[2]{#2}
\providecommand{\BIBentrySTDinterwordspacing}{\spaceskip=0pt\relax}
\providecommand{\BIBentryALTinterwordstretchfactor}{4}
\providecommand{\BIBentryALTinterwordspacing}{\spaceskip=\fontdimen2\font plus
\BIBentryALTinterwordstretchfactor\fontdimen3\font minus
  \fontdimen4\font\relax}
\providecommand{\BIBforeignlanguage}[2]{{%
\expandafter\ifx\csname l@#1\endcsname\relax
\typeout{** WARNING: IEEEtran.bst: No hyphenation pattern has been}%
\typeout{** loaded for the language `#1'. Using the pattern for}%
\typeout{** the default language instead.}%
\else
\language=\csname l@#1\endcsname
\fi
#2}}
\providecommand{\BIBdecl}{\relax}
\BIBdecl

\bibitem{smith2026datacenter}
S.~J. Smith \emph{et~al.}, ``{United States} data center energy usage report:
  2025 update,'' Lawrence Berkeley National Laboratory, Tech. Rep.
  LBNL-2001758, 2026.

\bibitem{colangelo2025phoenix}
\BIBentryALTinterwordspacing
P.~Colangelo \emph{et~al.}, ``Turning {AI} data centers into grid-interactive
  assets: Results from a field demonstration in {Phoenix, Arizona},'' 2025,
  arXiv:2507.00909.
\BIBentrySTDinterwordspacing

\bibitem{mohan2021checkfreq}
\BIBentryALTinterwordspacing
J.~Mohan, A.~Phanishayee, and V.~Chidambaram, ``{CheckFreq}: Frequent,
  fine-grained {DNN} checkpointing,'' in \emph{Proc. USENIX FAST}, 2021, pp. 203--216.
\BIBentrySTDinterwordspacing

\bibitem{lian2025universal}
\BIBentryALTinterwordspacing
X.~Lian \emph{et~al.}, ``Universal checkpointing: A flexible and efficient distributed
  checkpointing system for large-scale {DNN} training with reconfigurable
  parallelism,'' in \emph{Proc. USENIX ATC},
  2025, pp. 1519--1534.
\BIBentrySTDinterwordspacing

\bibitem{tsiligkaridis2025distributed}
A.~Tsiligkaridis, P.~Andrianesis, A.~K. Coskun, M.~C. Caramanis, and I.~C. Paschalidis, ``Distributed economic dispatch in power networks incorporating
  data center flexibility,'' \emph{IEEE Trans. Sustain. Comput.},
  vol.~10, no.~4, pp. 768--783, 2025.

\bibitem{zhang2024aggregator}
Y.~Zhang, A.~Tsiligkaridis, I.~C. Paschalidis, and A.~K. Coskun, ``Data center
  and load aggregator coordination towards electricity demand response,''
  \emph{Sustain. Comput. Inform. Syst.}, vol.~42, 2024, Art.~no.~100957.

\bibitem{tong2026execution}
\BIBentryALTinterwordspacing
X.~Tong, E.~Du, H.~Jiang, L.~Chen, and Q.~Wang, ``Execution-feasible dispatch
  of {LLM} pre-training loads for renewable-dominant {AI} data centers,'' 2026, arXiv:2610.13047.
\BIBentrySTDinterwordspacing

\bibitem{liu2026privacycheckpoint}
\BIBentryALTinterwordspacing
Z.~Liu, R.~Yang, X.~Cui, and F.~Y. Hou, ``Privacy-preserving coordinated
  operation of power grids and {AI} data centers: A checkpoint-aware
  three-phase scheme,'' 2026, arXiv:2609.26365.
\BIBentrySTDinterwordspacing

\bibitem{liu2011greening}
Z.~Liu, M.~Lin, A.~Wierman, S.~H. Low, and L.~L.~H. Andrew, ``Greening
  geographical load balancing,'' \emph{ACM SIGMETRICS Perform. Eval. Rev.}, vol.~39, no.~1, pp. 193--204, 2011.

\bibitem{hanafy2023carbonscaler}
W.~A. Hanafy, Q.~Liang, N.~Bashir, D.~Irwin, and P.~Shenoy, ``{CarbonScaler}:
  Leveraging cloud workload elasticity for optimizing carbon-efficiency,''
  \emph{Proc. ACM Meas. Anal. Comput. Syst.}, vol.~7, no.~3, 2023, Art.~no.~57.

\bibitem{radovanovic2023carbon}
A.~Radovanovi{\'c} \emph{et~al.}, ``Carbon-aware computing for datacenters,''
  \emph{IEEE Trans. Power Syst.}, vol.~38, no.~2, pp. 1270--1280,
  2023.

\bibitem{caprara2026flexibility}
A.~Caprara, Y.~Yu, F.~Teng, A.~Junyent-Ferr{\'e}, E.~Bullich-Massagu{\'e}, and M.~Arag{\"u}{\'e}s-Pe{\~n}alba, ``Data center workload flexibility for power
  system demand response: Evidence from {Alibaba} traces,'' \emph{Int. J. Electr. Power Energy Syst.}, vol.~178, 2026, Art.~no.~111940.

\bibitem{acun2026flexdc}
F.~Acun, C.~Hankendi, E.~Levine, H.~Reynolds, J.~Bardwick, and A.~K. Coskun,
  ``Investigating power consumption flexibility of {AI} data centers for demand
  response participation,'' in \emph{Proc. ACM e-Energy}, 2026,
  pp. 344--348.

\bibitem{norris2025rethinking}
\BIBentryALTinterwordspacing
T.~H. Norris, T.~Profeta, D.~Patino-Echeverri, and A.~Cowie-Haskell,
  ``Rethinking load growth: Assessing the potential for integration of large
  flexible loads in {US} power systems,'' Nicholas Institute for Energy,
  Environment \& Sustainability, Duke University, Durham, NC, Tech. Rep.
  NI R 25-01, Feb.
  2025.
\BIBentrySTDinterwordspacing

\bibitem{tsg2021aggregatedIDC}
M.~Chen \emph{et~al.},
  ``Aggregated model of data network for the provision of demand response in
  generation and transmission expansion planning,'' \emph{IEEE Trans. Smart Grid}, vol.~12, no.~1, pp. 512--523, 2021.

\bibitem{tsg2021incentiveIDC}
M.~Chen, C.~Gao, M.~Shahidehpour, and Z.~Li, ``Incentive-compatible demand
  response for spatially coupled internet data centers in electricity
  markets,'' \emph{IEEE Trans. Smart Grid}, vol.~12, no.~4, pp.
  3056--3069, 2021.

\bibitem{zheng2022alpa}
\BIBentryALTinterwordspacing
L.~Zheng \emph{et~al.}, ``{Alpa}: Automating
  inter- and intra-operator parallelism for distributed deep learning,'' in
  \emph{Proc. USENIX OSDI}, 2022, pp. 559--578.
\BIBentrySTDinterwordspacing

\bibitem{thorpe2023bamboo}
\BIBentryALTinterwordspacing
J.~Thorpe \emph{et~al.}, ``{Bamboo}: Making preemptible instances resilient for affordable
  training of large {DNNs},'' in \emph{Proc. USENIX NSDI}, 2023, pp. 497--513.
\BIBentrySTDinterwordspacing

\bibitem{vercellino2026genai}
\BIBentryALTinterwordspacing
R.~Vercellino \emph{et~al.}, ``Measurement of generative {AI} workload power
  profiles for whole-facility data center infrastructure planning,'' 2026,
  arXiv:2604.07345.
\BIBentrySTDinterwordspacing

\bibitem{choukse2025power}
\BIBentryALTinterwordspacing
E.~Choukse \emph{et~al.}, ``Power stabilization for {AI} training
  datacenters,'' 2025, arXiv:2508.14318.
\BIBentrySTDinterwordspacing

\bibitem{hu2024acmetrace}
Q.~Hu \emph{et~al.}, ``Characterization of large language model development in
  the datacenter,'' in \emph{Proc. USENIX NSDI}, 2024, pp. 709--729.

\bibitem{wan2025bytecheckpoint}
B.~Wan \emph{et~al.}, ``{ByteCheckpoint}: A unified checkpointing system
  for large foundation model development,'' in \emph{Proc. USENIX NSDI},
  2025, pp. 559--578.

\bibitem{lu2026battery}
X.~Lu, J.~Qiu, J.~Lin, S.~An, M.~Sun, and J.~Zhao, ``Battery-assisted operation of hyperscale {AI} data
centers under connect-and-manage interconnection practices,'' 2026,
arXiv:2605.14105.

\bibitem{liu2027interruptible}
B.~Liu, Q.~Guo, and R.-G. Cong, ``Interruptible {AI} training as virtual energy
  storage for renewable integration in computing centers,'' \emph{Energy for
  Sustainable Development}, vol.~96, 2027, Art.~no.~102159.

\bibitem{zhang2026carbon}
J.~Zhang and Y.~Wang, ``Sustainable grid integration of {AI} data centers:
Carbon-aware co-optimization of training workloads and battery storage under
renewable uncertainty,'' \emph{Sustainability}, vol.~18, no.~18, 2026, Art.~no.~9504.

\bibitem{doe2024datacenter}
\BIBentryALTinterwordspacing
O.~Van Geet and D.~Sickinger, ``Best practices guide for energy-efficient data
  center design,'' National Renewable Energy Laboratory, Tech. Rep.
  NREL/TP-5R00-89843, 2024.
\BIBentrySTDinterwordspacing

\bibitem{stott2009dc}
B.~Stott, J.~Jardim, and O.~Alsa{\c{c}}, ``{DC} power flow revisited,''
  \emph{IEEE Trans. Power Syst.}, vol.~24, no.~3, pp. 1290--1300,
  2009.

\bibitem{desrosiers2005primer}
J.~Desrosiers and M.~E. L{\"u}bbecke, ``A primer in column generation,'' in
  \emph{Column Generation}, G.~Desaulniers, J.~Desrosiers, and
  M.~M.~Solomon, Eds. Boston, MA, USA: Springer, 2005, pp. 1--32.

\bibitem{barnhart1998branch}
C.~Barnhart, E.~L. Johnson, G.~L. Nemhauser, M.~W.~P. Savelsbergh, and P.~H. Vance, ``Branch-and-price: Column generation for solving huge integer
  programs,'' \emph{Oper. Res.}, vol.~46, no.~3, pp. 316--329, 1998.

\bibitem{wentges1997weighted}
P.~Wentges, ``Weighted {Dantzig--Wolfe} decomposition for linear mixed-integer
  programming,'' \emph{Int. Trans. Oper. Res.},
  vol.~4, no.~2, pp. 151--162, 1997.

\bibitem{barrows2020rts}
C.~Barrows \emph{et~al.}, ``The {IEEE} reliability test system: A proposed 2019
  update,'' \emph{IEEE Trans. Power Syst.}, vol.~35, no.~1, pp.
  119--127, 2020.

\bibitem{pglib2023}
IEEE PES Task Force on Benchmarks for Validation of Emerging Power System
  Algorithms, \emph{Power Grid Library (PGLib-OPF)}, ver.~23.07, 2023.

\bibitem{pasqualin2024async}
L.~Pasqualin \emph{et~al.}, ``Reducing model checkpointing times by over 10x
  with PyTorch distributed asynchronous checkpointing,'' PyTorch, Jun. 2024.

\bibitem{nyiso_realtime}
New York Independent System Operator, \emph{Real-Time Market LBMP -- Zonal (P-24A)}, 2024--2025.

\end{thebibliography}
\end{document}